# Homogeneous-isotropic sector of loop quantum gravity: new approach

### Marcin Kisielowski

National Centre for Nuclear Research, Pateura 7, 02-093 Warsaw, Poland

E-mail: Marcin.Kisielowski@ncbj.gov.pl

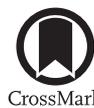

**Abstract**
Recently, a new class of scalar constraint operators has been introduced in loop quantum gravity. They are defined on a space of solutions to the Gauss constraint and partial solutions to the vector constraint, called a vertex Hilbert space. We propose a subspace of the vertex Hilbert space formed by homogeneous-isotropic states, which is invariant under the action of the new scalar constraint operators. As a result, the operators can be reduced to our homogeneous-isotropic subspace. The (generalized) eigenstates of the reduced operator are eigenstates of the full operator. We discuss the feasibility of numerical diagonalization of the reduced scalar constraint operator.

## 1. Introduction

In Loop Quantum Gravity substantial work has been done to construct a theory of quantum gravity in mathematically rigorous way [1–7]. Promising candidates have been proposed [8, 9] and deriving physical predictions seems to be of only technical nature.

Homogeneous-isotropic sector of the theory has been studied first by performing symmetry reduction classically and quantizing the resulting theory [10–13]. This procedure leads to a theory called loop quantum cosmology, which predicts that big-bang singularity is replaced by big-bounce due to quantum gravity effects. Later, the symmetry reduction has been also studied at the quantum level: using coherent states [14–23] or imposing some conditions on the states [24, 25]. One of the important unsolved problems is to find homogeneous-isotropic semiclassical states in full LQG preserved by the evolution generated by the physical Hamiltonian.

We will propose new approach to symmetry reduction: we will construct subspaces of states which are homogeneous-isotropic and are invariant under the action of the physical Hamiltonian. As a result, the operator descends to the symmetry-reduced subspaces. The eigenstates of the reduced operator are eigenstates of the full scalar constraint operator. This opens the possibility of repeating the steps from LQC [13] and verifying if our approach predicts big-bounce scenario. In particular the homogeneous-isotropic semiclassical states in full LQG preserved by evolution can be constructed by complete analogy from LQC. The only problem is to diagonalize the physical Hamiltonian. We plan to make the computations numerically. In this paper we will introduce a cut-off in the Hilbert space, which is compatible with our symmetry reduction and we will estimate the range of the cut-off, where numerical diagonalization of the physical Hamiltonian is viable.

Our construction applies to a class of physical Hamiltonians introduced recently [26–29]. Important examples of such operators appear in two LQG models: gravity coupled to massless scalar field and irrotational dust [29]. The physical Hamiltonian is obtained after deparametrization with respect to the scalar field or dust field [9, 27, 29–33]. In both cases it is a function of the gravitational part of the scalar constraint operator. Therefore we will focus on the diagonalization of this operator, which we will call in this paper simply the scalar constraint operator.

## 2. Vertex Hilbert space

The new operators [26–29] are defined on the vertex Hilbert space. For detailed description of this space we refer our reader to [26]. In this section we recall the definition and an isomorphism induced by the Peter–Weyl theorem. This section is just a compilation of some basic constructions in loop quantum gravity (see for example [1]) with the definition of the vertex Hilbert space [26]. Its purpose is to discuss in detail an isomorphism which will be used in the following.

### 2.1. Kinematical Hilbert space

The kinematical Hilbert space of loop quantum gravity $\mathcal{H}_{\text{kin}}$ is built from cylindrical functions of SU(2) connections on spatial manifold $\Sigma$. A function $\Psi$ is called cylindrical with respect to a graph $\gamma$ embedded in $\Sigma$ if there is suitably regular function $\psi : SU(2)^N \to \mathbb{C}$ such that

$$\Psi(A) = \psi(A_{\ell_1}, \ldots, A_{\ell_N}), \tag{1}$$

where $\ell_1, \ldots, \ell_N$ are the links of the graph $\gamma$ and $A_\ell$ is a holonomy of the connection $A$ along the link $\ell$. For precise definition of an embedded graph we refer our reader to [26] (in [26] links are called edges and nodes are called vertices). In our paper we will denote by Links($\gamma$) the set of the links of the graph $\gamma$ and by Nodes($\gamma$) the set of the nodes of the graph $\gamma$:

$$\text{Links}(\gamma) = \{\ell_1, \ldots, \ell_N\}, \quad \text{Nodes}(\gamma) = \{n_1, n_2, \ldots, n_M\}. \tag{2}$$

By suitably regular function $\psi(u_1, \ldots, u_N)$ we mean polynomial in $\rho_1(u_1), \ldots, \rho_N(u_N)$, where $\rho_I$ are SU(2) representations including the trivial one. A function that is cylindrical with respect to some graph is simply called cylindrical. A function is cylindrical with respect to many graphs, for example a function cylindrical with respect to a graph $\gamma$ is cylindrical with respect to any graph $\gamma'$ obtain from $\gamma$ by a sequence of operations of flipping orientations of links, splitting links and adding links [1, 2, 34–36]. Given two graphs $\gamma$ and $\gamma'$ (and assuming a suitable differentiability class of the links [37]) it is always possible to find a graph $\gamma''$ which is obtained by a sequence of such operations from $\gamma$ and $\gamma'$. As a result, two functions $\Psi$ and $\Psi'$ cylindrical with respect to graphs $\gamma$ and $\gamma'$, respectively, are also cylindrical with respect to $\gamma''$.

Using this property, it is possible to define a scalar product in the space of cylindrical functions:

$$\langle \Psi | \Psi' \rangle = \int du_1 \ldots du_{N''} \overline{\psi}(u_1, \ldots, u_{N''}) \psi'(u_1, \ldots, u_{N''}), \quad (3)$$

where $N''$ is the number of links in $\gamma''$. The kinematical Hilbert space is the Cauchy completion of the space of cylindrical functions with respect to the scalar product (3):

$$\mathcal{H}_{\text{kin}} := \overline{\text{Cyl}}. \quad (4)$$

The space of functions cylindrical with respect to a graph $\gamma$ will be denoted by $\text{Cyl}_\gamma$. The Cauchy completion of $\text{Cyl}_\gamma$ will be denoted by $\tilde{\mathcal{H}}_\gamma$:

$$\tilde{\mathcal{H}}_\gamma := \overline{\text{Cyl}_\gamma}. \quad (5)$$

The space $\tilde{\mathcal{H}}_\gamma$ is isomorphic to $L^2(G^N)$. By the Peter–Weyl theorem this space is isomorphic to

$$\tilde{\mathcal{H}}_\gamma \cong \bigotimes_{\ell \in \text{Links}(\gamma)} \bigoplus_{j_\ell} S_{j_\ell}, \quad (6)$$

where

$$S_j = \mathcal{H}_j \otimes \mathcal{H}_j^* \quad (7)$$

is the tensor product of the SU(2) representation space $\mathcal{H}_j$ with spin $j$ and its dual $\mathcal{H}_j^*$. The (inverse) isomorphism will be denoted by $\varphi_\gamma$:

$$\varphi_\gamma : \bigotimes_{\ell \in \text{Links}(\gamma)} \bigoplus_{j_\ell} S_{j_\ell} \to \tilde{\mathcal{H}}_\gamma. \quad (8)$$

If $\gamma$ is a graph obtained from $\gamma'$ by a sequence of operations of splitting links and adding links, then

$$\tilde{\mathcal{H}}_{\gamma'} < \tilde{\mathcal{H}}_\gamma. \quad (9)$$

Following [26] we will say that $\Psi \in \tilde{\mathcal{H}}_\gamma$ is a proper element of $\tilde{\mathcal{H}}_\gamma$ if the following holds:

$$\tilde{\mathcal{H}}_{\gamma'} < \tilde{\mathcal{H}}_\gamma \Rightarrow \Psi \perp \tilde{\mathcal{H}}_{\gamma'}. \quad (10)$$

The subspace of $\tilde{\mathcal{H}}_\gamma$ formed by proper states will be denoted by $\mathcal{H}_\gamma$. The kinematical Hilbert space is a direct sum (of orthogonal Hilbert spaces):

$$\mathcal{H}_{\text{kin}} = \bigoplus_\gamma \mathcal{H}_\gamma. \quad (11)$$

### 2.2. The space of solutions to the Gauss constraint

The Gauss constraint generates SU(2) gauge transformations. Every such gauge transformation $g : \Sigma \to \text{SU}(2)$ induces a transformation of the cylindrical functions

$$\Psi(A) \mapsto \Psi(gAg^{-1} + g^{-1}dg). \quad (12)$$

The space of solutions to the Gauss constraint $\mathcal{H}^{\text{Gauss}}$ is the Cauchy completion of the space $\text{Cyl}^{\text{Gauss}}$ of cylindrical functions invariant under the action of all SU(2) gauge transformations:

$$\mathcal{H}^{\text{Gauss}} = \overline{\text{Cyl}^{\text{Gauss}}}. \tag{13}$$

The completion of the subspace of SU(2) gauge invariant functions cylindrical with respect to a graph $\gamma$ will be denoted by $\tilde{\mathcal{H}}^{\text{Gauss}}_\gamma$:

$$\tilde{\mathcal{H}}^{\text{Gauss}}_\gamma := \overline{\text{Cyl}_\gamma \cap \text{Cyl}^{\text{Gauss}}}. \tag{14}$$

The subspace of $\tilde{\mathcal{H}}^{\text{Gauss}}_\gamma$ formed by proper states will be denoted by $\mathcal{H}^{\text{Gauss}}_\gamma$. The space $\mathcal{H}^{\text{Gauss}}$ has the following orthogonal decomposition:

$$\mathcal{H}^{\text{Gauss}} = \bigoplus_\gamma \mathcal{H}^{\text{Gauss}}_\gamma. \tag{15}$$

In order to find a useful basis of $\mathcal{H}^{\text{Gauss}}$, let us write the decomposition

$$\tilde{\mathcal{H}}_\gamma \cong \bigotimes_{\ell \in \text{Links}(\gamma)} \bigoplus_{j_\ell} S_{j_\ell}, \tag{16}$$

in an equivalent form

$$\bigotimes_{\ell \in \text{Links}(\gamma)} \bigoplus_{j_\ell} S_{j_\ell} = \bigoplus_{j_\ell} \bigotimes_{n \in \text{Nodes}(\gamma)} \mathcal{H}_n, \tag{17}$$

where

$$\mathcal{H}_n = \mathcal{H}^*_{j_1} \otimes \ldots \otimes \mathcal{H}^*_{j_M} \otimes \mathcal{H}_{j_{M+1}} \otimes \ldots \otimes \mathcal{H}_{j_N}, \tag{18}$$

links $\ell_1, \ldots, \ell_M$ are incoming to the node $n$ and links $\ell_{M+1}, \ldots, \ell_N$ are outgoing. The subspace $\tilde{\mathcal{H}}^{\text{Gauss}}_\gamma$ is obtained by restricting at each node $n$ to the space $\text{Inv}(\mathcal{H}_n)$ of tensors invariant under the action of the SU(2) group, i.e.

$$\tilde{\mathcal{H}}^{\text{Gauss}}_\gamma = \varphi_\gamma \left( \bigoplus_{j_\ell} \bigotimes_{n \in \text{Nodes}(\gamma)} \text{Inv}(\mathcal{H}_n) \right). \tag{19}$$

A convenient basis of $\mathcal{H}^{\text{Gauss}}$ is labelled by the spin networks. A spin network $s$ is a triple $(\gamma, \rho, \iota)$:

- an oriented graph $\gamma$ (an embedded graph with a choice of orientation of each link),
- labelling $\rho$ of the links of the graph with unitary irreducible representations of the SU(2) group:

$$\rho_\ell \in \text{Irrep}(\text{SU}(2)), \tag{20}$$

- labelling $\iota$ of the nodes with tensors invariant under the action of the group

$$\iota_n \in \text{Inv}\left( \mathcal{H}^*_{\rho_{\ell_1}} \otimes \ldots \otimes \mathcal{H}^*_{\rho_{\ell_M}} \otimes \mathcal{H}_{\rho_{\ell_{M+1}}} \otimes \ldots \otimes \mathcal{H}_{\rho_{\ell_N}} \right), \tag{21}$$

where links $\ell_1, \ldots, \ell_M$ are incoming to the node $n$ and links $\ell_{M+1}, \ldots, \ell_N$ are outgoing.

Since the dimension of the unitary irreducible representation $\rho_\ell$ defines an integer or half-integer $j_\ell$ by the formula $\dim \rho_\ell = 2j_\ell + 1$, it is clear that the spin networks defined on a graph $\gamma$ label a basis in $\bigoplus_{j_\ell} \bigotimes_{n \in \text{Nodes}(\gamma)} \text{Inv}(\mathcal{H}_n)$ and therefore in $\tilde{\mathcal{H}}_\gamma^{\text{Gauss}}$.

### 2.3. The space of solutions to the vector constraint and the vertex Hilbert space

The vector constraint generates diffeomorphisms of $\Sigma$. Every analytic diffeomorphism $f \in \text{Diff}(\Sigma)$ defines a unitary operator $U_f : \mathcal{H}_{\text{kin}} \to \mathcal{H}_{\text{kin}}$:

$$U_f \Psi(A) = \Psi(f^*A). \tag{22}$$

In [26] the authors propose to impose the quantum vector constraint in two steps. The first step is to average the states in each space $\mathcal{H}_\gamma$ with respect to all diffeomorphisms $\text{Diff}_{\text{Nodes}(\gamma)}$ that act trivially on the set $\text{Nodes}(\gamma)$ of nodes of the graph $\gamma$. This procedure leads to the vertex Hilbert space. The second step is to average with respect to the remaining diffeomorphisms $\text{Diff}(\Sigma)/\text{Diff}_{\text{Nodes}(\gamma)}$.

The first averaging is done by considering a map $\eta$

$$\mathcal{H}_\gamma \ni \Psi \mapsto \eta(\Psi) \in \text{Cyl}^* \tag{23}$$

defined by

$$\eta(\Psi) := \frac{1}{N_\gamma} \sum_{[f] \in \text{Diff}_{\text{Nodes}(\gamma)}/\text{TDiff}_\gamma} < U_f \Psi |. \tag{24}$$

In the formula above $\text{TDiff}_\gamma$ is formed by diffeomorphisms acting trivially on $\mathcal{H}_\gamma$. In the averaging procedure the quotient is used to avoid infinite factors coming from averaging with respect to diffeomorphisms which have trivial action on the state. The factor $N_\gamma$ can be arbitrary but it is convenient to set it equal to order of the group

$$\text{Sym}_\gamma = \{f \in \text{Diff}_{\text{Nodes}(\gamma)} : f(\gamma) = \gamma\}/\text{TDiff}_\gamma \tag{25}$$

of symmetries of the graph $\gamma$ fixing each node of the graph:

$$N_\gamma = \#\text{Sym}_\gamma. \tag{26}$$

The state $\eta(\Psi)$ is invariant under the action of the diffeomorphisms $\text{Diff}_{\text{Nodes}(\gamma)}$ that have trivial action on the nodes of the graph $\gamma$. It is therefore partial solution to the vector constraint. If two graphs $\gamma$ and $\gamma'$ can be related by an action of a diffeomorphism fixing each element in the set $\text{Nodes}(\gamma) = \text{Nodes}(\gamma')$, then the spaces $\eta(\mathcal{H}_\gamma)$ and $\eta(\mathcal{H}_{\gamma'})$ are the same:

$$\eta(\mathcal{H}_\gamma) = \eta(\mathcal{H}_{\gamma'}). \tag{27}$$

It is therefore justified to introduce a notation:

$$\mathcal{H}_{[\gamma]} = \eta(\mathcal{H}_\gamma), \tag{28}$$

where $[\gamma]$ is $\text{Diff}_{\text{Nodes}(\gamma)}$-equivalence class of graphs with representative $\gamma$. Each space $\mathcal{H}_{[\gamma]}$ is isometric to $S_\gamma$—the subspace of $\mathcal{H}_\gamma$ of elements invariant under the action of the group $\text{Sym}_\gamma$. With the choice of the factor $N_\gamma = \#\text{Sym}_\gamma$, the isometry is given by the map $\eta$:

$$\eta : S_\gamma \to \eta(\mathcal{H}_\gamma). \tag{29}$$

The subspaces of $\mathcal{H}_{[\gamma]}$ and $S_\gamma$ formed by states invariant under SU(2) gauge transformations will be denoted by $\mathcal{H}_{[\gamma]}^{\text{Gauss}}$ and $S_\gamma^{\text{Gauss}}$, respectively:

$$\mathcal{H}_{[\gamma]}^{\text{Gauss}} := \eta(\mathcal{H}_\gamma^{\text{Gauss}}), \quad S_\gamma^{\text{Gauss}} := S_\gamma \cap \mathcal{H}_\gamma^{\text{Gauss}}. \tag{30}$$

The map $\eta$ can be extended by linearity to the orthogonal sum:

$$\eta : \bigoplus_\gamma \mathcal{H}_\gamma \to \text{Cyl}^*.$$

Since $\text{Cyl} \subset \mathcal{H}_\gamma$, the space of cylindrical functions is in the domain of the map. In the original definition [26] the vertex Hilbert space is a completion of $\eta(\text{Cyl})$ under the norm defined by the scalar product

$$\langle \eta(\Psi) | \eta(\Psi') \rangle := \eta(\Psi)(\Psi'). \tag{31}$$

In the following we will restrict to SU(2) gauge invariant cylindrical functions $\text{Cyl}^{\text{Gauss}}$ and we will define the vertex Hilbert space as

$$\mathcal{H}_{\text{vtx}} := \overline{\eta(\text{Cyl}^{\text{Gauss}})}. \tag{32}$$

Our definition coincides with the definition used for example in [27].

The space $\mathcal{H}_{\text{vtx}}$ has an orthogonal decomposition:

$$\mathcal{H}_{\text{vtx}} = \bigoplus_{V \in \text{FS}(\Sigma)} \mathcal{H}_V, \tag{33}$$

where $\text{FS}(\Sigma)$ is the set of finite subsets of $\Sigma$. Each state in $\mathcal{H}_V$ is invariant under the action of the group $\text{Diff}_V$ of diffeomorphism that act trivially on the set $V \in \text{FS}(\Sigma)$. Each space $\mathcal{H}_V$ has an orthogonal decomposition:

$$\mathcal{H}_V = \bigoplus_{[\gamma] \in [\gamma(V)]} \mathcal{H}_{[\gamma]}^{\text{Gauss}}, \tag{34}$$

where

$$[\gamma(V)] := \{[\gamma] : \text{Nodes}(\gamma) = V\}. \tag{35}$$

The states in $\mathcal{H}_V$ are invariant under $\text{Diff}_V$. In order to obtain states invariant under action of all diffeomorphisms, second averaging needs to be performed. It is sufficient to consider states in $\mathcal{H}_{\text{vtx}}$ of the form $\eta(\Psi)$, where $\Psi \in \mathcal{H}_\gamma$. The state $\eta(\Psi)$ is mapped to Diff-invariant state in the following way:

$$\mathcal{H}_{\text{vtx}} \ni \eta(\Psi) \mapsto \sum_{[f] \in \text{Diff}/\text{Diff}_{\text{Nodes}(\gamma)}} \eta(U_f \Psi) \in \text{Cyl}^*. \tag{36}$$

## 3. Scalar constraint operators

In the following we will focus on particular class of scalar constraint operators. We will describe their general features in this section. To our knowledge there are two examples of such operators studied in [29]. As in the aforementioned paper we will be interested in diagonalizing the operators.

### 3.1. General properties of the scalar constraint operator

We assume that the scalar constraint operator $\hat{C}(N)$ is densely defined on $\mathcal{H}_{\text{vtx}}$ and has the following properties:

(a) It is local:
$$\hat{C}(N) = \sum_{x \in \Sigma} N(x)\hat{C}_x, \tag{37}$$

$$\hat{C}_x \mathcal{H}_V \subset \mathcal{H}_V, \tag{38}$$

$$\hat{C}_x|_{\mathcal{H}_V} = 0, \quad \text{unless } x \in V \tag{39}$$

(b) It is covariant:
$$U_f \hat{C}_x U_f^{-1} = \hat{C}_{f(x)} \quad \text{for all } f \in \text{Diff}(\Sigma). \tag{40}$$

(c) Each operator $\hat{C}_x$ has the following splitting
$$\hat{C}_x = \sum_{\ell, \ell'} \epsilon(\dot{\ell}, \dot{\ell}') \hat{C}_{x\,\ell,\ell'}, \tag{41}$$

where the sum is over pairs of links at the node $x$. The coefficient $\epsilon(\dot{\ell}, \dot{\ell}')$ is zero if the vectors $\dot{\ell}, \dot{\ell}'$ are colinear and 1 otherwise. Each of the operators $\hat{C}_{x\ell,\ell'}$ satisfies:
$$\hat{C}_{x\,\ell,\ell'} \mathcal{H}_{[\gamma]} \subset \mathcal{H}_{[\gamma_-]} \oplus \mathcal{H}_{[\gamma]} \oplus \mathcal{H}_{[\gamma_+]}, \tag{42}$$

where $\gamma_+$ is obtained from $\gamma$ by adding a loop tangential to links $\ell, \ell'$ according to the prescription defined in the next subsection and $\gamma_-$ is obtained by removing such loop.

(d) The operator does not change representation labels of the links. The new loop or removed loop is labelled with fixed unitary irreducible representation $\rho_{(l)}$ of dimension $2l + 1$. The corresponding representation space will be denoted by $\mathcal{H}_{(l)}$.

(e) Each operator $\hat{C}_x$ does not change the intertwiners associated to nodes different from $x$.

See also [26].

### 3.2. A note about regularization of the operator

We will consider the constraint operators defined in [27–29]. In the definition of the operator there is some freedom: the so-called Euclidean part adds or subtracts a loop tangential to two different links of the graphs but different types of loops lead to different operators. We will consider the following definition. Important role in the definition is played by the concept of tangentiality order of links. Two curves $c_1 : [0, 1] \to \Sigma$ and $c_2 : [0, 1] \to \Sigma$ have contact order $k$, if for every smooth function $F : \Sigma \to \mathbb{R}$ the difference $F \circ c_1 - F \circ c_2$ and all its (right) derivatives up to order $k$ vanish at $0 \in [0, 1]$ (see also for example the beginning of chapter 12 in [38]). The tangentiality order of links $\ell, \ell'$ will be the highest possible contact order of curves obtained by parametrizing the links $\ell, \ell'$. Let us notice that the order of tangentiality of links is diffeomorphism invariant. This has the following important consequence. Let us consider two graphs $\gamma_1$ and $\gamma_2$ depicted on figure 1. It is clear that the Hilbert spaces $\mathcal{H}_{\gamma_1}$ and $\mathcal{H}_{\gamma_2}$ are orthogonal. Due to the diffeomorphism invariance of the tangentiality order of links, the Hilbert spaces $\mathcal{H}_{[\gamma_1]}$ and $\mathcal{H}_{[\gamma_2]}$ are also orthogonal. In a sense, the tangentiality order of links is one of the properties of embedded graphs that still plays a role after averaging with respect to the diffeomorphisms. We will say that a loop is between links $\ell_I$ and $\ell_J$ if and only if it is tangential to links $\ell_I$ and $\ell_J$. We will assume that a loop between links $\ell_I$ and $\ell_J$ added by the Euclidean part will have the same order of tangentiality with both links, denoted by $T_{IJ}$. It will be called the order of the loop. The highest order of tangentiality of loops between links $\ell_I$ and $\ell_J$ will be called an order of the wedge $(\ell_I, \ell_J)$ and denoted $\mathfrak{l}(I, J)$. When a loop is added by the Euclidean part we assume that its order of tangentiality is equal to $\mathfrak{l}(I, J) + 1$. When a loop is subtracted by the Euclidean part, it will be assumed that the loop with the highest order is removed. This regularization makes the loops distinguishable but we do not keep track of the sequence in which the loops have been added.

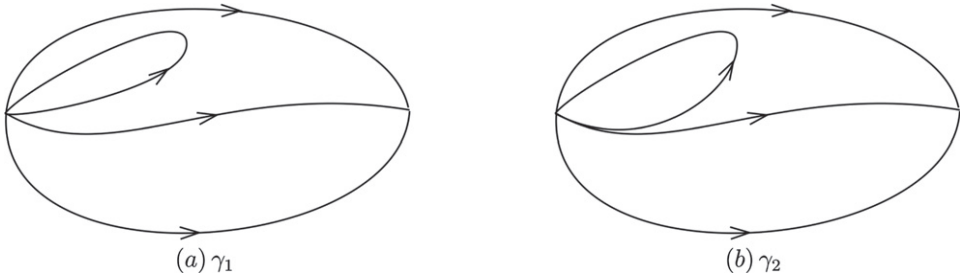

**Figure 1.** An example illustrating the role of the tangentiality order of links in the vertex Hilbert space. We consider two different graphs $\gamma_1$ and $\gamma_2$ depicted on figures (a) and (b), respectively. It is clear that $\mathcal{H}_{\gamma_1} \perp \mathcal{H}_{\gamma_2}$. In graph $\gamma_2$ the loop is tangential to another link of the graph (tangentiality order 1) and the loop in graph $\gamma_1$ is not tangential to any other link of the graph (tangentiality order 0 with each link). Since tangentiality order is diffeomorphism invariant, the Hilbert spaces $\mathcal{H}_{[\gamma_1]}$ and $\mathcal{H}_{[\gamma_2]}$ are also orthogonal.

One could choose a different regularization. For example in [27] the authors propose the following prescription for adding loops. They introduce a concept of the order of tangentiality $k_I$ of the link $\ell_I$ at the node $n$, which is the highest order of tangentiality of the link $\ell_I$ with the remaining links incident at $n$. A loop $\alpha_{IJ}$ added by the Euclidean part has the order of tangentiality $k_I + 1$ and $k_J + 1$ with links $\ell_I$ and $\ell_J$, respectively.

In order to explain the motivation behind our regularization, we will consider a neighbourhood of a 3-valent node $n$ with pairwise non-tangential links each connecting two different nodes. Consider two sequences of adding loops depicted schematically on figure 2. Let us denote by $T(\alpha_{ij}, \ell_k)$ the tangentiality order of the loop $\alpha_{ij}$ with the link $\ell_k$. In our regularization the loops $\alpha_{12}$ and $\alpha_{13}$ are tangent to the corresponding links up to order 1, i.e. the non-zero numbers $T(\alpha_{ij}, \ell_k)$ are all equal 1:

$$T(\alpha_{13}, \ell_1) = T(\alpha_{13}, \ell_3) = T(\alpha_{12}, \ell_1) = T(\alpha_{12}, \ell_2) = 1. \tag{43}$$

In the regularization from [27] the tangentiality orders are different for the two sequences. For sequence (a) they are:

$$T(\alpha_{13}, \ell_1) = T(\alpha_{13}, \ell_3) = 1, \quad T(\alpha_{12}, \ell_1) = 2, \quad T(\alpha_{12}, \ell_2) = 1. \tag{44}$$

For sequence (b) they are:

$$T(\alpha_{13}, \ell_1) = 2, \quad T(\alpha_{13}, \ell_3) = 1, \quad T(\alpha_{12}, \ell_1) = T(\alpha_{12}, \ell_2) = 1. \tag{45}$$

We chose our prescription, because the truncated invariant subspaces that we will introduce in section 4.3 are smaller.

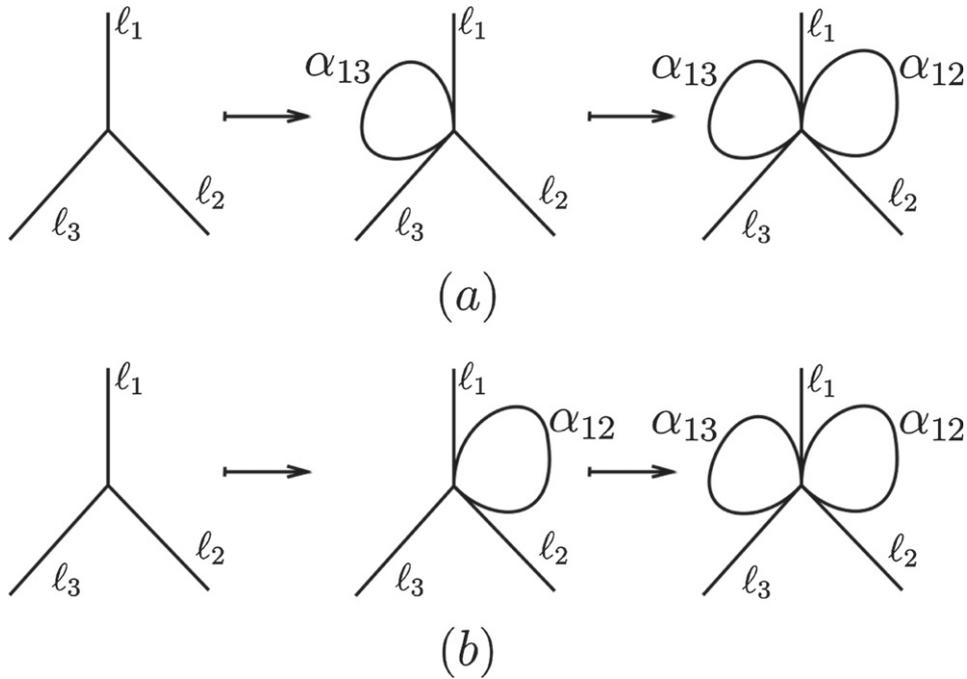

**Figure 2.** The difference between our regularization and the regularization from [27] is illustrated by considering two sequences of adding loops (a) and (b). In the first step of sequence (a) we add a loop $\alpha_{13}$ tangential to links $\ell_1$ and $\ell_3$, in the second step we add a loop $\alpha_{12}$ tangential to links $\ell_1$ and $\ell_2$. In sequence (b) we make the same steps but in reverse order. The result of the sequences of steps depends on the regularization. In our regularization the non-zero tangentiality orders $T(\alpha_{ij}, \ell_k)$ are all equal 1 (see (43)). In the regularization from [27] the tangentiality orders are different for sequence (a) and (b) (see (44) and (45)).

### 3.3. Invariant subspace

We will be interested in subspaces of the vertex Hilbert space which are invariant under the action of the operator $\hat{C}_x$. They are useful when looking for the (generalized) eigenvalues of the operator, because if a vector satisfies eigen-equation, it in particular means that it is in invariant subspace.

Firstly, let us notice that from property (a) from section 3.1 it follows that $\mathcal{H}_V$ is an invariant subspace. However, it can be further split into invariant subspaces. Let $\Gamma$ be a graph without loops (an example of such graph is the lattice from section 4.1). As a result of the assumption about the number of links connecting two different nodes, at each node $x$ we can choose ordering of the links $\ell_1, \ldots, \ell_{N_x}$. Let $\{\mathfrak{l}_n\}_{n \in \text{Nodes}(\Gamma)}$ be a family of maps

$$\mathfrak{l}_n : \{(I, J) : I < J, \epsilon(\dot{\ell}_I, \dot{\ell}_J) \neq 0\} \to \mathbb{N}_n. \tag{46}$$

Let $\Gamma_\mathfrak{l}$ be a graph obtained from $\Gamma$ by adding, at each node $n$, $\mathfrak{l}_n(I, J)$ loops tangent at $n$ to the links $\ell_I$ and $\ell_J$ such that the beginning of each loop is tangent to $\ell_I$, the end is tangent to $\ell_J$, and the orders of the loops are $1, \ldots, \mathfrak{l}_n(I, J)$. Clearly, from the property (c) it follows that the space

$$\mathcal{H}_\Gamma^{\text{loop}} := \bigoplus_\mathfrak{l} \mathcal{H}_{[\Gamma_\mathfrak{l}]}^{\text{Gauss}} \subset \mathcal{H}_V, \quad V = \text{Nodes}(\Gamma) \tag{47}$$

is also invariant (let us recall that in our notation $\mathcal{H}^{\text{Gauss}}_{[\Gamma_{\mathfrak{l}}]} = \eta(\mathcal{H}^{\text{Gauss}}_{\Gamma_{\mathfrak{l}}})$ and $[\Gamma_{\mathfrak{l}}]$ is the Diff$_V$-equivalence class of graphs). In order to simplify the further discussion, we will assume that the group of symmetries of $\Gamma$ fixing each node of $\Gamma$ is trivial #Sym$_\Gamma = 1$. In this case the averaging map

$$\eta : \mathcal{H}^{\text{Gauss}}_{\Gamma_{\mathfrak{l}}} \to \mathcal{H}^{\text{Gauss}}_{[\Gamma_{\mathfrak{l}}]} \tag{48}$$

is an isometry ($\mathcal{H}^{\text{Gauss}}_{\Gamma_{\mathfrak{l}}} = S^{\text{Gauss}}_{\Gamma_{\mathfrak{l}}}$). Each of the spaces $\mathcal{H}^{\text{Gauss}}_{\Gamma_{\mathfrak{l}}}$ can be further decomposed using Peter–Weyl theorem

$$\mathcal{H}^{\text{Gauss}}_{\Gamma_{\mathfrak{l}}} \cong \bigoplus_{j_\ell} \bigotimes_{n \in \text{Nodes}(\Gamma_{\mathfrak{l}})} \text{Inv}(\mathcal{H}_n). \tag{49}$$

Let us notice that for different numbers of loops $\mathfrak{L}(\mathfrak{l}_n)$ in $\mathfrak{l}_n$ the spaces $\text{Inv}(\mathcal{H}_n)$ are different. To make this dependence more explicit we introduce the following notation:

$$\mathcal{H}_{\mathfrak{L}(\mathfrak{l}_n)} = \text{Inv}\left(\underbrace{\mathcal{H}_{(l)} \otimes \mathcal{H}^*_{(l)} \otimes \ldots \otimes \mathcal{H}_{(l)} \otimes \mathcal{H}^*_{(l)}}_{2\mathfrak{L}(\mathfrak{l}_n)} \otimes \mathcal{H}_{j_1} \otimes \ldots \otimes \mathcal{H}_{j_N}\right), \tag{50}$$

where $\mathcal{H}_{(l)}$ denotes the spin $l$ representation space of SU(2): $\dim \mathcal{H}_{(l)} = 2l+1$. In summary, the invariant space $\mathcal{H}^{\text{loop}}_\Gamma$ has a decomposition

$$\mathcal{H}^{\text{loop}}_\Gamma = \bigoplus_{j_\ell} \mathcal{H}^{\text{loop}}_{\Gamma,j}, \tag{51}$$

where

$$\mathcal{H}^{\text{loop}}_{\Gamma,j} \cong \bigoplus_{\mathfrak{l}} \bigotimes_{n \in \text{Nodes}(\Gamma_{\mathfrak{l}})} \mathcal{H}_{\mathfrak{L}(\mathfrak{l}_n)}. \tag{52}$$

From property (d) if follows that each of the subspaces $\mathcal{H}^{\text{loop}}_{\Gamma,j}$, such that $j_\alpha = j_{(l)}$ for each loop $\alpha$, is invariant under the action the operator $\hat{C}_n$ for each node $n \in \text{Nodes}(\Gamma)$. From the locality property (a) it further follows that the space is actually invariant under the action of $\hat{C}(N)$.

Let us notice that our invariant space $\mathcal{H}^{\text{loop}}_{\Gamma,j}$ is isomorphic to

$$\mathcal{H}^{\text{loop}}_{\Gamma,j} \cong \bigotimes_{n \in \text{Nodes}(\Gamma_{\mathfrak{l}})} \bigoplus_{\mathfrak{l}_n} \mathcal{H}_{\mathfrak{L}(\mathfrak{l}_n)}. \tag{53}$$

From the property (e) it follows that each of the subspaces

$$\mathcal{H}^{\text{loop}}_{\Gamma,j,x} := \bigoplus_{\mathfrak{l}_x} \mathcal{H}_{\mathfrak{L}(\mathfrak{l}_x)} \tag{54}$$

is invariant under the action of the operator $\hat{C}_x$. In the diagonalization problem it is enough to find the eigenvalues of the operators $\hat{C}_x$ separately. By taking tensor product of eigenstates corresponding to different $x$ we obtain simultaneous eigenstates of the operators $\hat{C}_x$, where $x \in \text{Nodes}(\Gamma)$.

Let us notice, that we have used so far all properties except for the covariance property (b). It will play central role in the symmetry reduction proposed in the next section.

## 4. The symmetric states

### 4.1. Pure lattice

*4.1.1. Symmetries.* In this paper our spatial manifold will be the 3-torus:
$$\Sigma = \mathbb{T}^3. \tag{55}$$

It will be convenient to consider it as a quotient space [39]. Given a real number $a \in \mathbb{R}$, let us consider a subgroup of the group of translations $\mathcal{T}_a = \{T_{(pa,qa,ra)} : p, q, r \in \mathbb{Z}\}$, where $T_v$ denotes a translation by a vector $v$. The quotient space $\mathbb{R}^3/\mathcal{T}_a$ is diffeomorphic to $\mathbb{T}^3$:
$$\mathbb{T}^3 = \mathbb{R}^3/\mathcal{T}_a. \tag{56}$$

Let us fix a number $\epsilon \in \mathbb{R}$ such that $\frac{1}{\epsilon} \in \mathbb{N}^+$. Let us consider an infinite cubical lattice $\widetilde{\Gamma}$ in $\mathbb{R}^3$ such that:

- the coordinates of the nodes are
$$(p\epsilon a, q\epsilon a, r\epsilon a), \tag{57}$$
  where $p, q, r \in \mathbb{Z}$,
- there is precisely 1 link connecting each nearest-neighbour pair of nodes.

This lattice naturally descends to a lattice on $\mathbb{T}^3 = \mathbb{R}^3/\mathcal{T}_a$. We will denote the corresponding lattice on $\mathbb{T}^3$ by $\Gamma$.

We consider a group $O_{\text{cube}}$ of orientation preserving symmetries of a cube. This group is a subgroup of a group generated by matrices
$$R_0 = \begin{pmatrix} 1 & 0 & 0 \\ 0 & 1 & 0 \\ 0 & 0 & -1 \end{pmatrix}, \quad R_1 = \begin{pmatrix} 1 & 0 & 0 \\ 0 & 0 & 1 \\ 0 & 1 & 0 \end{pmatrix}, \quad R_2 = \begin{pmatrix} 0 & 1 & 0 \\ 1 & 0 & 0 \\ 0 & 0 & 1 \end{pmatrix} \tag{58}$$

formed by matrices of determinant 1. The subgroup of the Euclidean group $\mathcal{T}_{\epsilon a} \rtimes O_{\text{cube}}$ acting in $\mathbb{R}^3$ preserves the infinite lattice $\widetilde{\Gamma}$. Let us denote by $\mathcal{T}_\epsilon$ the quotient group:
$$\mathcal{T}_\epsilon := \mathcal{T}_{\epsilon a}/\mathcal{T}_a. \tag{59}$$

The group
$$\mathcal{T}_\epsilon \rtimes O_{\text{cube}} \tag{60}$$

is a subgroup of isometries of $\mathbb{T}^3 = \mathbb{R}^3/\mathcal{T}_a$ equipped with the canonical flat metric (the metric for which $\pi : \mathbb{R}^3 \to \mathbb{R}^3/\mathcal{T}_a$ is the Riemannian covering map [39]). The group $\mathcal{T}_\epsilon \rtimes O_{\text{cube}}$ preserves the lattice $\Gamma$.

*4.1.2. Homogeneous isotropic states.* We will say that a spin network $s$ is homogeneous-isotropic spin network if is invariant under $\mathcal{T}_\epsilon \rtimes O_{\text{cube}}$ up to a phase, i.e. for any diffeomorphism $g \in \mathcal{T}_\epsilon \rtimes O_{\text{cube}}$
$$g^*|s> = e^{i\Phi_s(g)}|s>, \tag{61}$$

where $e^{i\Phi_s(g)}$ is one-dimensional unitary representation of the symmetry group $\mathcal{T}_\epsilon \rtimes O_{\text{cube}}$. For homogeneous isotropic spin networks it follows in particular that
$$\forall_{\ell \in \text{Links}(\gamma)} \dim \rho_\ell = 2j + 1 \tag{62}$$

for a given fixed spin $j$. The phase, which we will choose in the following section, depends only on this spin $j$. Therefore, instead of writing $\Phi_s(g)$ we will write $\Phi_j(g)$.

Let us focus on the intertwiner space corresponding to a single node $n$. Let us denote the links incident at $n$ by $\ell_1, \ell_2, \ldots, \ell_6$. Without loss of generality, we can assume that the coordinates of the node are $(0,0,0)$ and that all links are outgoing from the node. To each group element $g \in O_\text{cube}$ there corresponds a permutation of the links incident at the node $n$. We will denote the permutation corresponding to $g$ by $\sigma_g$. We consider a projection operator $P_j : \text{Inv}\left((\mathcal{H}_j)^{\otimes 6}\right) \to \text{Inv}\left((\mathcal{H}_j)^{\otimes 6}\right)$ acting on an intertwiner $\iota$ in the following way:

$$(P_j \iota)^{A_1 \ldots A_6} = \frac{1}{24} \sum_{g \in O_\text{cube}} e^{-i\Phi_j(g)} \iota^{A_{\sigma_g^{-1}(1)} \ldots A_{\sigma_g^{-1}(6)}}. \tag{63}$$

The image of $\text{Inv}\left((\mathcal{H}_j)^{\otimes 6}\right)$ with the projection operator $P_j$ will be denoted by $\mathcal{H}_{j,0}^\text{cube}$ (the meaning of the index 0 will be explained in section 4.3):

$$\mathcal{H}_{j,0}^\text{cube} := P_j(\text{Inv}\left((\mathcal{H}_j)^{\otimes 6}\right)). \tag{64}$$

For isotropic states all nodes are labelled with the same (or equivalent) intertwiner, an element of $\mathcal{H}_{j,0}^\text{cube}$. Our states are similar to states used in GFT condensate approach [24, 25]. However, instead of restricting to highest volume eigenvalue in the invariant subspace, we restrict to purely symmetric states.

*4.1.3. Fixing the phase.* Let us notice that the group $O_\text{cube}$ is isomorphic to the group $S_4$ of permutations of the 4 diagonals of the cube. To each element $g \in O_\text{cube}$ we will assign a number $\text{sgn}(g) = \pm 1$ equal to the sign of the corresponding permutation $\nu_g$ in $S_4$:

$$\text{sgn}(g) := \text{sgn}(\nu_g). \tag{65}$$

We choose the phase to be equal to

$$e^{i\Phi_j(g)} = (\text{sgn}(g))^{2j}, \tag{66}$$

where $j$ is the spin assigned to each link. With this choice of phase the space $\mathcal{H}_{j,0}^\text{cube}$ has at least two non-trivial elements. Both are the Livine–Speziale coherent intertwiners [40]:

$$|j, \vec{n}_1 \ldots \vec{n}_6 > := \int_{SU(2)} du \bigotimes_{i=1}^{6} \rho_j(u) |j, \vec{n}_i >. \tag{67}$$

First corresponds to a cube oriented in accordance with standard orientation of $\mathbb{R}^3$:

$$\vec{n}_1 = \begin{pmatrix} 1 \\ 0 \\ 0 \end{pmatrix}, \quad \vec{n}_2 = \begin{pmatrix} -1 \\ 0 \\ 0 \end{pmatrix}, \quad \vec{n}_3 = \begin{pmatrix} 0 \\ 1 \\ 0 \end{pmatrix}, \quad \vec{n}_4 = \begin{pmatrix} 0 \\ -1 \\ 0 \end{pmatrix}, \quad \vec{n}_5 = \begin{pmatrix} 0 \\ 0 \\ 1 \end{pmatrix}, \quad \vec{n}_6 = \begin{pmatrix} 0 \\ 0 \\ -1 \end{pmatrix}. \tag{68}$$

Second corresponds to a cube with opposite orientation:

$$\vec{n}_1 = \begin{pmatrix} -1 \\ 0 \\ 0 \end{pmatrix}, \quad \vec{n}_2 = \begin{pmatrix} 1 \\ 0 \\ 0 \end{pmatrix}, \quad \vec{n}_3 = \begin{pmatrix} 0 \\ 1 \\ 0 \end{pmatrix}, \quad \vec{n}_4 = \begin{pmatrix} 0 \\ -1 \\ 0 \end{pmatrix}, \quad \vec{n}_5 = \begin{pmatrix} 0 \\ 0 \\ 1 \end{pmatrix}, \quad \vec{n}_6 = \begin{pmatrix} 0 \\ 0 \\ -1 \end{pmatrix}. \tag{69}$$

The phase factor is calculated in the following way. The Perelomov coherent states transform under the action of the SU(2) group element $u$ according to the following formula [41]:

$$\rho_j(u)|j, \vec{n}\rangle = \exp(i\Phi(\vec{n}, u))|j, u \cdot \vec{n}\rangle, \tag{70}$$

where $u \cdot \vec{n}$ is the unit vector obtained by rotating $\vec{n}$ with the SO(3) matrix corresponding to $u \in \mathrm{SU}(2)$,

$$\Phi(\vec{n}, u) = jA(\vec{n}_0, \vec{n}, u \cdot \vec{n}), \tag{71}$$

$A(\vec{n}_0, \vec{n}, u \cdot \vec{n})$ is the area of the geodesical triangle on the sphere, with the vertices at the points $\vec{n}_0 = (0, 0, 1)^\mathrm{T}, \vec{n}, u \cdot \vec{n}$. The action of the group element $g \in O_{\mathrm{cube}}$ on $\mathbb{R}^3$ leads to the permutation $\sigma_g$ of the vectors $\vec{n}_i$:

$$g \cdot \vec{n}_i = \vec{n}_{\sigma_g^{-1}(i)}. \tag{72}$$

From invariance of the Haar measure it follows that

$$|j, \vec{n}_{\sigma_g^{-1}(1)} \ldots \vec{n}_{\sigma_g^{-1}(6)}\rangle = |j, g \cdot \vec{n}_1 \ldots g \cdot \vec{n}_6\rangle \sim |j, \vec{n}_1 \ldots \vec{n}_6\rangle. \tag{73}$$

Using formula (70) we obtained that the phase factor relating $|j, g \cdot \vec{n}_1 \ldots g \cdot \vec{n}_6\rangle$ and $|j, \vec{n}_1 \ldots \vec{n}_6\rangle$ is equal to (66):

$$|j, g \cdot \vec{n}_1 \ldots g \cdot \vec{n}_6\rangle = (\mathrm{sgn}(g))^{2j}|j, \vec{n}_1 \ldots \vec{n}_6\rangle. \tag{74}$$

Since $g$ is orientation preserving, it is possible to obtain any permutation of the normals $\vec{n}_1, \ldots, \vec{n}_6$ corresponding to rotations of the cube. It is not possible for example to obtain a permutation corresponding to reflection by the $y - z$ plane. Therefore the second vector configuration (69) is in principle not related to the first configuration (68). In fact, direct calculation shows that for $j = \frac{1}{2}$ the Livine–Speziale coherent intertwiners corresponding to (68) and (69) are linearly independent (but not orthogonal).

*4.1.4. Symmetry reduction in the space of invariants.* Thanks to the choice of phase, the image of the projection is always non-trivial. In fact our numerical calculations indicate that for large spins the dimension of the space grows polynomially (see figure 3(a)). Let us introduce a ratio

$$r(j) = \frac{\dim \mathrm{Inv}\left((\mathcal{H}_j)^{\otimes 6}\right)}{\dim \mathcal{H}_{j,0}^{\mathrm{cube}}}. \tag{75}$$

Our numerical study indicates that this ratio grows with increasing spin (see figure 3(b)). Unfortunately, at this point, we are not able to see if the values of the ratio are bounded or not. Of particular interest for the study performed in section 5.2 is the question if the value of $r(j)$ can exceed 24. For this we would need to do calculations for larger spins, which would require more advanced parallelization of our code. Since this is not the main objective of this paper, we leave it to future research.

Interestingly, let us notice that

$$\dim \mathcal{H}_{\frac{1}{2},0}^{\mathrm{cube}} = 2. \tag{76}$$

Direct calculation shows that the two Livine–Speziale coherent intertwiners corresponding to the two cubes with opposite orientations span this space.

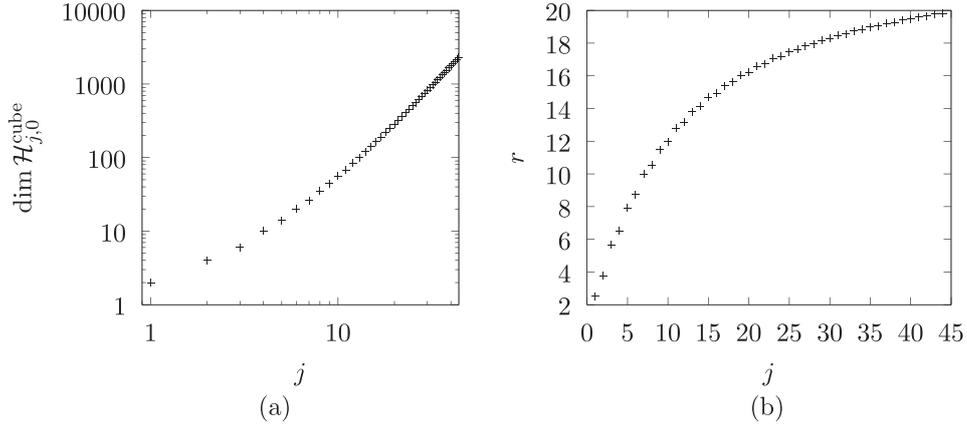

**Figure 3.** The dimension of the symmetry reduced space $\mathcal{H}_{j,0}^{\text{cube}}$ (a) and the ratio $r$ of the dimension of the full invariant space to the dimension of the symmetry reduced space (b) for different values of spin $j$. On figure (a) logarithmic scale on $x$ and $y$-axis is used.

### 4.2. Lattice with loops

Let us consider a diffeomorphism $f_g$ corresponding to an element $g$ in $O_{\text{cube}}$. The diffeomorphism leaves the space $\mathcal{H}_{\Gamma,j,x}^{\text{loop}}$ invariant. Since we assumed that $\hat{C}_x$ is covariant with respect to diffeomorphisms (see property (b) in section 3.1), the operator

$$P_j^{\text{cube}} = \frac{1}{24} \sum_{g \in O_{\text{cube}}} e^{-i\Phi_j(g)} U_{f_g} \tag{77}$$

commutes with the operator $\hat{C}_x$:

$$P_j^{\text{cube}} \hat{C}_x = \hat{C}_x P_j^{\text{cube}}. \tag{78}$$

As a result the space $\mathcal{H}_j^{\text{cube}} = \text{Im}(P_j^{\text{cube}})$ is also invariant under the action of $\hat{C}_x$ (let us notice that there is no conflict of notation with section 4.1 because the operator $P_j^{\text{cube}}$ in section 4.1 is just a restriction of the operator defined above). Let us denote by $\mathcal{H}_j^{\text{sym}}$ the diagonal subspace of $\bigotimes_{n \in \text{Nodes}(\Gamma)} \bigoplus_{l_n} \mathcal{H}_{\mathfrak{L}(l_n)}$ corresponding to $\mathcal{H}_j^{\text{cube}}$, i.e. a subspace spanned by vectors

$$\bigotimes_{n \in \text{Nodes}(\Gamma)} v_n, \quad \text{where } v_n = v \in \mathcal{H}_j^{\text{cube}}. \tag{79}$$

The states are invariant (up to a phase) under the action of the symmetry group $\mathcal{T}_\epsilon \rtimes O_{\text{cube}}$. We will also denote by

$$\mathcal{H}^{\text{sym}} = \bigoplus_j \mathcal{H}_j^{\text{sym}}. \tag{80}$$

### 4.3. Truncation

Our goal will be to look for the eigenstates numerically. To this end we will truncate the invariant Hilbert space $\mathcal{H}_{\Gamma,j,x,L}^{\text{loop}}$ to loop configurations $\mathfrak{l}$ with at most $L$ loops ($\mathfrak{L}(\mathfrak{l}) < L$), i.e. we will

consider a space

$$\mathcal{H}^{\text{loop}}_{\Gamma,j,x,L} = \bigoplus_{\mathfrak{l}_x : \mathfrak{L}(\mathfrak{l}_x) \leqslant L} \mathcal{H}_{\mathfrak{L}(\mathfrak{l}_x)}. \tag{81}$$

Let us notice that this truncation is compatible with the symmetry reduction. This means that the projection operator $P_j^{\text{cube}}$ leaves the subspace $\mathcal{H}^{\text{loop}}_{\Gamma,j,x,L}$ invariant. As a result we introduce symmetry reduced truncated space:

$$\mathcal{H}^{\text{cube}}_{j,L} = P_j^{\text{cube}}(\mathcal{H}^{\text{loop}}_{\Gamma,j,x,L}). \tag{82}$$

## 5. Symmetry reduction of the invariant space

In the previous section we introduced an subspace invariant under the action of operator $\hat{C}_x$ and introduced a cut-off which makes numerical calculations possible. In this section we will study a reduction of the truncated space due to averaging with respect to orientation preserving symmetries of a cube $O_{\text{cube}}$. The action of the diffeomorphisms $f_g$ corresponding to $g \in O_{\text{cube}}$ on the spin-network states induces an action in $\mathcal{H}^{\text{loop}}_{\Gamma,j,x} = \bigoplus_{\mathfrak{l}_x} \mathcal{H}_{\mathfrak{L}(\mathfrak{l}_x)}$:

$$U_{f_g}(\mathfrak{l}_x, \iota_x) = (g \cdot \mathfrak{l}_x, R(g)\iota_x). \tag{83}$$

The action of $g$ on $\mathfrak{l}$ is the following:

$$(g \cdot \mathfrak{l})(I,J) = \begin{cases} \mathfrak{l}(g^{-1}(I), g^{-1}(J)), & \text{if } g^{-1}(I) < g^{-1}(J) \\ \mathfrak{l}(g^{-1}(J), g^{-1}(I)) & \text{otherwise.} \end{cases} \tag{84}$$

The action of $g$ on the intertwiners is more complicated and we will devote to it the next subsection.

### 5.1. The action of the group of cubical symmetries on the intertwiner spaces

#### 5.1.1. Assigning indices to links.
Let us describe how the indices of the intertwiners are assigned to the links. We focus on the node $x$ with loop configuration $\mathfrak{l}$. Let us recall that by $\mathfrak{L}(\mathfrak{l})$ we denoted the total number of loops:

$$\mathfrak{L}(\mathfrak{l}) = \sum_{I<J} \mathfrak{l}(I,J). \tag{85}$$

A tensor in the space

$$\mathcal{H}_{(l)} \otimes \mathcal{H}^*_{(l)} \otimes \ldots \otimes \mathcal{H}_{(l)} \otimes \mathcal{H}^*_{(l)} \otimes \mathcal{H}_{j_1} \otimes \ldots \otimes \mathcal{H}_{j_6} \tag{86}$$

has $2\mathfrak{L}(\mathfrak{l}) + 6$ indices. Let us call the indices $A_1, \ldots, A_{2\mathfrak{L}(\mathfrak{l})+6}$. Indices $A_1, \ldots, A_{2\mathfrak{L}(\mathfrak{l})}$ correspond to the loops and $A_{2\mathfrak{L}(\mathfrak{l})+1}, \ldots, A_{2\mathfrak{L}(\mathfrak{l})+6}$ correspond to the remaining links (dual to faces of a cube). We can label each loop by indices $(I, J, \mathfrak{l}(I,J) - T_{IJ} + 1)$ (let us recall that $T_{IJ}$ is the order of the loop). We will use lexicographical order to compare the triples $(I, J, K)$ and $(I', J', K')$. Let us consider a sequence of the indices

$$(\beta_1, \ldots, \beta_{\mathfrak{L}(\mathfrak{l})}) \tag{87}$$

where $\beta_i < \beta_j \iff i < j$. We will assign to each loop another label: a number $i \in \{1, \ldots, \mathfrak{L}(\mathfrak{l})\}$ such that $\beta_i$ is the labelling of the loop. To the loop $\beta_i$ there correspond two indices $A_{2i}$ and $A_{2i+1}$. The index $A_{2i}$ is up and the index $A_{2i+1}$ is down.

*5.1.2. The action of the group $O_{\text{cube}}$ on the indices.*   To each group element $g \in O_{\text{cube}}$ there corresponds a diffeomorphism which induces a permutation $\sigma_g \in S_{2\mathfrak{L}(\mathfrak{l})+6}$ of the indices of the intertwiner.

There is a natural action of the group $O_{\text{cube}}$ on the sequence

$$(1, 2, \ldots, 2\mathfrak{L}(\mathfrak{l}) + 6). \tag{88}$$

The group acts on the indices $(I, J, K)$:

$$g \cdot (I, J, K) = \begin{cases} (g(I), g(J), K), & \text{if } g(I) < g(J) \\ (g(J), g(I), K), & \text{otherwise} \end{cases} \tag{89}$$

In the second case $(g \cdot (I, J, K) = (g(J), g(I), K))$ we will say that $g$ flips the loop $(I, J, K)$. This action induces a permutation $\mu_g \in S_{\mathfrak{L}(\mathfrak{l})}$ in the following way. We will consider a sequence $\beta^g$ obtained from a set $\{g \cdot \beta_i : i \in \{1, \ldots, \mathfrak{L}(\mathfrak{l})\}\}$ by ordering its elements lexicographically. The permutation $\mu_g$ is defined in the following way:

$$\beta^g_{\mu_g(i)} = g \cdot \beta_i. \tag{90}$$

As a result the action of $g$ induces a permutation $\sigma_g \in S_{2\mathfrak{L}(\mathfrak{l})+6}$

$$\sigma_g(i) = \begin{cases} 2\mu_g(\frac{i+1}{2}), & \text{if } i <= 2\mathfrak{L}(\mathfrak{l}) \text{ and } i \text{ is odd} \\ & \text{and } g \text{ does not flip the loop } \beta_{\frac{i+1}{2}}, \\ 2\mu_g(\frac{i}{2}) + 1, & \text{if } i <= 2\mathfrak{L}(\mathfrak{l}) \text{ and } i \text{ is even} \\ & \text{and } g \text{ does not flip the loop } \beta_{\frac{i}{2}}, \\ 2\mu_g(\frac{i+1}{2}) + 1, & \text{if } i <= 2\mathfrak{L}(\mathfrak{l}) \text{ and } i \text{ is odd and } g \text{ flips the loop } \beta_{\frac{i+1}{2}}, \\ 2\mu_g(\frac{i}{2}), & \text{if } i <= 2\mathfrak{L}(\mathfrak{l}) \text{ and } i \text{ is even and } g \text{ flips the loop } \beta_{\frac{i}{2}}, \\ g(i - 2\mathfrak{L}(\mathfrak{l})) + 2\mathfrak{L}(\mathfrak{l}), & \text{if } i > 2\mathfrak{L}(\mathfrak{l}). \end{cases} \tag{91}$$

One could think at first sight that the action of the group $O_{\text{cube}}$ on the intertwiner is defined by the permutation $\sigma_g$ in completely analogous way as in (63). This is in fact the case when $g$ interpreted as a diffeomorphism acting on a spin network does not flip orientation of any loop. If an orientation of a loop is flipped by the diffeomorphism, we need to use equivalent spin network where the orientation of the loop is flipped back to the proper orientation. However, the equivalence move leads to non-trivial action on the space of intertwiners, which we will discuss below.

*5.1.3. The representation of the group $O_{\text{cube}}$ in the intertwiner space.*   It is well known fact that for each of the SU(2) group representations there is a bilinear form invariant under the action of the group:

$$\rho_j(u)^A_B \rho_j(u)^C_D \epsilon_{jAC} = \epsilon_{jBD}. \tag{92}$$

If the dimension of the representation is even, the form is antisymmetric, and if it is odd, the form is symmetric:

$$\epsilon_{jBA} = (-1)^{2j} \epsilon_{jAB}. \tag{93}$$

The form defines an intertwiner $\epsilon : \mathcal{H}_j \to \mathcal{H}_j^*$ between the representation $\rho_j$ and its dual representation $\rho_j^*$:

$$\rho_j^*(u)\epsilon_j = \epsilon_j \rho_j(u). \tag{94}$$

In the index notation:

$$\rho_j^*(u)_A{}^B \epsilon_{jBC} = \epsilon_{jAD} \rho_j(u)^D{}_C \tag{95}$$

The equation (93) can be written as

$$\epsilon_j^* = (-1)^{2j} \epsilon_j. \tag{96}$$

We will use the form $\epsilon$ to define the action of a permutation $\sigma \in S_N$ on $N$-valent intertwiners:

$$\mathrm{Inv}\left(\mathcal{H}_{j_1}^* \otimes \ldots \otimes \mathcal{H}_{j_K}^* \otimes \mathcal{H}_{j_{K+1}} \otimes \ldots \otimes \mathcal{H}_{j_N}\right). \tag{97}$$

Firstly, we define the action of the permutation on an intertwiner in $\mathrm{Inv}\left(\mathcal{H}_{j_1} \otimes \ldots \otimes \mathcal{H}_{j_N}\right)$:

$$(\sigma \cdot \iota)^{A_1 \ldots A_N} = \iota^{A_{\sigma^{-1}(1)} \ldots A_{\sigma^{-1}(N)}}. \tag{98}$$

Let us notice that

$$\begin{aligned}\epsilon_{j_1} \otimes \ldots \otimes \epsilon_{j_K} \otimes \mathbb{1} \otimes \ldots \otimes \mathbb{1} : & \mathrm{Inv}\left(\mathcal{H}_{j_1} \otimes \ldots \otimes \mathcal{H}_{j_N}\right) \\ & \to \mathrm{Inv}\left(\mathcal{H}_{j_1}^* \otimes \ldots \otimes \mathcal{H}_{j_K}^* \otimes \mathcal{H}_{j_{K+1}} \otimes \ldots \otimes \mathcal{H}_{j_N}\right)\end{aligned} \tag{99}$$

is an isomorphism. We define the action of the permutation $\sigma$ on

$$\iota \in \mathrm{Inv}\left(\mathcal{H}_{j_1}^* \otimes \ldots \otimes \mathcal{H}_{j_K}^* \otimes \mathcal{H}_{j_{K+1}} \otimes \ldots \otimes \mathcal{H}_{j_N}\right) \tag{100}$$

in the following way:

$$\sigma \cdot \iota := \left(\eta_{j_1} \otimes \ldots \otimes \eta_{j_N}\right) \cdot \sigma \cdot \left(\epsilon_{j_1} \otimes \ldots \otimes \epsilon_{j_K} \otimes \mathbb{1} \otimes \ldots \otimes \mathbb{1}\right)^{-1} \cdot \iota, \tag{101}$$

where

$$\eta_{j_I} = \begin{cases} \epsilon_{j_{\sigma^{-1}(I)}}, & \text{if } \sigma^{-1}(I) \leqslant K \\ \mathbb{1} & \text{otherwise.} \end{cases} \tag{102}$$

If a diffeomorphism $g$ flips a loop $\beta_i$, then the corresponding permutation acts between spaces:

$$\sigma_g : \mathrm{Inv}\left(\ldots \otimes \mathcal{H}_{(l)} \otimes \mathcal{H}_{(l)}^* \otimes \ldots\right) \to \mathrm{Inv}\left(\ldots \otimes \mathcal{H}_{(l)}^* \otimes \mathcal{H}_{(l)} \otimes \ldots\right). \tag{103}$$

This reflects the fact that in the corresponding spin network $s$ the loop has opposite orientation to the standard one. However, this spin network is equivalent to the spin network $s'$ with a loop in the standard orientation but in $s'$ this loop is labelled with representation $\rho_{(l)}^*$. As we mentioned at the beginning of the section, the representation $\rho_{(l)}^*$ is equivalent to the representation $\rho_{(l)}$:

$$\rho_{(l)}^*(u) = \epsilon_{(l)}\, \rho_{(l)}(u)\, \epsilon_{(l)}^{-1}. \tag{104}$$

This means that the spin network $s'$ is equivalent to spin network $s''$ where the loop has the standard orientation and it is labelled with representation $\rho_l$ but the node $x$ is labelled with an invariant tensor

$$\iota''_x = \mathbb{1} \otimes \ldots \otimes \mathbb{1} \otimes \epsilon_{(l)}^{-1} \otimes \epsilon_{(l)}^* \otimes \mathbb{1} \otimes \ldots \otimes \mathbb{1}\, \iota_x. \tag{105}$$

Using the property (96) we find that:

$$\iota''_x = (-1)^{2l} \mathbb{1} \otimes \ldots \otimes \mathbb{1} \otimes \epsilon^{-1}_{(l)} \otimes \epsilon_{(l)} \otimes \mathbb{1} \otimes \ldots \otimes \mathbb{1} \, \iota_x. \tag{106}$$

Let us denote by $F(g, \mathfrak{l})$ the (minimal) number of flips required to bring all loops in the spin network $s$ to the standard orientation (the beginning of a loop is tangent to $\ell_I$ and the end of a loop is tangent to $\ell_J$, where $I < J$).

As a result, the group $O_{\text{cube}}$ has the following representation $R(g)$:

$$R(g) : \text{Inv}\left(\mathcal{H}_{(l)} \otimes \mathcal{H}^*_{(l)} \otimes \ldots \otimes \mathcal{H}_{(l)} \otimes \mathcal{H}^*_{(l)} \otimes \mathcal{H}_{j_1} \otimes \ldots \otimes \mathcal{H}_{j_6}\right)$$
$$\to \text{Inv}\left(\mathcal{H}_{(l)} \otimes \mathcal{H}^*_{(l)} \otimes \ldots \otimes \mathcal{H}_{(l)} \otimes \mathcal{H}^*_{(l)} \otimes \mathcal{H}_{j_1} \otimes \ldots \otimes \mathcal{H}_{j_6}\right), \tag{107}$$

$$R(g)\iota = (-1)^{2l\,F(g,\mathfrak{l})} \mathbb{1} \otimes \epsilon_{(l)} \otimes \ldots \otimes \mathbb{1} \otimes \epsilon_{(l)} \otimes \mathbb{1} \otimes \ldots \otimes \mathbb{1} \cdot \sigma_g \cdot \mathbb{1} \otimes$$
$$\epsilon^{-1}_{(l)} \otimes \ldots \otimes \mathbb{1} \otimes \epsilon^{-1}_{(l)} \otimes \mathbb{1} \otimes \ldots \otimes \mathbb{1} \cdot \iota. \tag{108}$$

### 5.2. The outlook for numerical calculations

Knowing the action of $O_{\text{cube}}$ on $\mathcal{H}^{\text{loop}}_{\Gamma,j,x}$ we can write the action of the projection operator $P^{\text{cube}}_j$ (77) on a vector $(\mathfrak{l}, \iota) \in \mathcal{H}^{\text{loop}}_{\Gamma,j,x}$:

$$P^{\text{cube}}_j(\mathfrak{l}, \iota) = \frac{1}{24} \sum_{g \in O_{\text{cube}}} e^{-i\Phi_j(g)} (g \cdot \mathfrak{l}, R(g)\iota). \tag{109}$$

This formula allows practical calculations. The rate of symmetry reduction $r(j, L)$ will be defined to be the ratio of the dimensions of the spaces $\mathcal{H}^{j,L}_{\text{loops}}$ and $\mathcal{H}^{j,L}_{\text{cube}}$.

$$r(j,L) = \frac{\dim \mathcal{H}^{\text{loop}}_{\Gamma,j,x,L}}{\dim \mathcal{H}^{\text{cube}}_{j,L}} = \frac{\sum_{\mathfrak{l}_x : \mathfrak{L}(\mathfrak{l}_x) \leqslant L} \dim \mathcal{H}_{\mathfrak{L}(\mathfrak{l}_{\tilde{x}})}}{\text{Tr}\, P^{\text{cube}}_{j,L}}, \tag{110}$$

where $P^{\text{cube}}_{j,L}$ denotes the restriction of $P^{\text{cube}}_j$ to the subspace $\mathcal{H}^{\text{loop}}_{\Gamma,j,x,L}$. Let us notice that for cubical lattice there are three pairs of links which are colinear. This means that instead of considering in $\mathfrak{l}_x$ 15 possible pairs of links, we consider only 12 pairs of links such that $\epsilon(\dot{\ell}_I, \dot{\ell}_J) \neq 0$.

On figures 4 and 5 we present the results of our numerical calculations. On figure 4 the dimension of the space $\mathcal{H}^{\text{cube}}_{j,L}$ is plotted in logarithmic scale. The plot indicates exponential growth of the dimension of the matrix with increasing truncation $L$. We plotted the lowest spins $j \in \{\frac{1}{2}, 1, \frac{3}{2}, \ldots, 5\}$. The matrices with rank up to $10^6$ can be typically diagonalized on a supercomputer. Therefore the plot shows which values of parameters $j$ and $L$ are within reach of numerical study. On figure 5 the ratio $r(j, L)$ is plotted as a function of $L$. It grows with increasing $L$ and stabilizes fast at value close to 24. It is not clear for us if the behaviour is the same for all spins. In particular, as we mentioned in section 4.1.4 we do not know if $r(j) = r(j, 0)$ has values always smaller than 24. If $r(j, 0)$ exceeds the value 24 for some spins, the behaviour of $L \mapsto r(j, L)$ for such spins must be different from the one observed in this paper. The rate of symmetry reduction around 24 obtained for the range of spins studied in this paper is substantial. Since diagonalization algorithms have complexity roughly $\mathcal{O}(n^3)$, where $n$ is the rank of the matrix, we obtain a speedup of around $10^4$ times compared to naive approach without projection in the invariant subspaces.

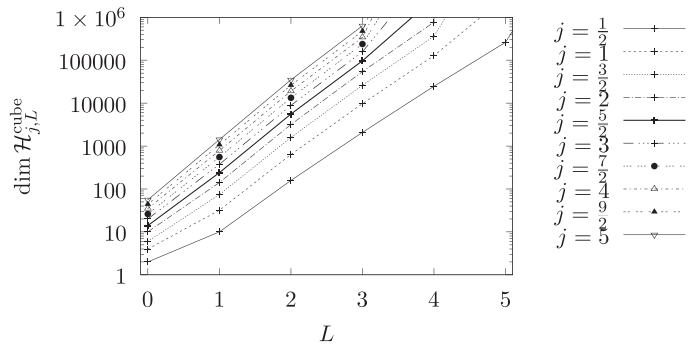

**Figure 4.** The dimension of the symmetry reduced space truncated up to $L$ loops plotted for different values of spin $j$. Logarithmic scale on $y$-axis is used. The region of $\dim \mathcal{H}^{j,L}_{\text{cube}} < 10^6$ is plotted. The matrices with rank up to $10^6$ can be typically diagonalized on a supercomputer.

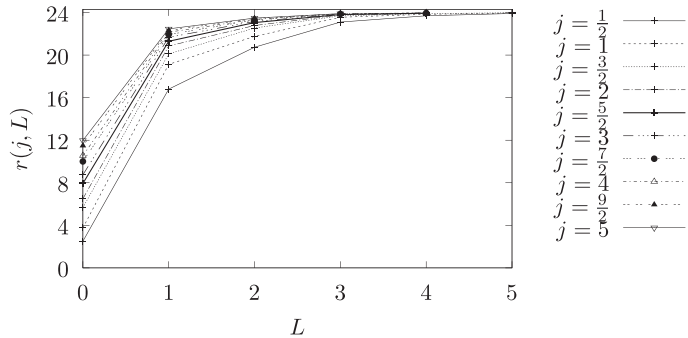

**Figure 5.** The ratio of the dimension of the full invariant space truncated up to $L$ loops to the dimension of the symmetry reduced space.

## 6. Summary and discussion

The restriction to homogeneous-isotropic sector of loop quantum gravity that we proposed leads to substantial reduction of the degrees of freedom. From the plethora of spin-network states defined on cubical lattice $\Gamma$, it restricts us to lattices with links labelled with the same spin. Moreover, it allows us to restrict the problem of diagonalization of operators $\hat{C}_n$, $n \in \text{Nodes}(\Gamma)$ to single operator $\hat{C}_x$ at a fixed node $x$. The symmetry group acts non-trivially in the intertwiner spaces, leading, after averaging, to further reduction of the degrees of freedom. We have noticed that the averaging should include non-trivial phase factor to accommodate Livine–Speziale coherent intertwiners. We found, that after a truncation of the relevant Hilbert space to spin networks with not more than $L$ loops at each node, the symmetry reduction leads to almost 24 times smaller subspaces of intertwiners.

Let us notice, that similar restriction on the intertwiner spaces has been considered in [24, 25]. However, in [24, 25] each intertwiner space is reduced to one-dimensional subspace spanned by the highest volume eigenvalue. However, this condition is not preserved by the scalar constraint.

Our restriction is not so drastic as in [24, 25] but the factor 24 is substantial. Most diagonalization algorithms have complexity close to $\mathcal{O}(n^3)$, where $n$ is the rank of the matrix. This means around $10^4$ speedup compared to naive approach. The computing centre were our numerical calculation were done has $3 \times 10^4$ cores. This roughly speaking means that instead of running a diagonalization programme using all resources of our computing centre, we could just run the programme on our laptop. Practically, this means that instead of considering truncation with $L$ loops we can consider truncation with $L + 1$ loops. This may become crucial for deriving some physically interesting results.

We hope that this speedup will allow us to find eigenstates of the operators $\hat{C}_x$. We realize that the truncation needs further study. We plan to look for its justification using numerical calculations: by varying the truncation and investigating if some of the eigenstates converge when the number of loops increases. Such calculations could also give some insight on analytic properties of the truncation.

The resulting eigenstates states could be in principle compared with eigenstates of loop quantum cosmology. To this end we would expand the eigenstates of the truncated full theory in the eigenbasis of the volume operator and compare them as functions of the corresponding eigenvalues. The eigenstates in the region of large volumes could be compared with the Wheeler–DeWitt eigenstates in the homogeneous-isotropic sector. This could also be a method to verify the classical limit of the Hamiltonians proposed in [27–29].

Let us also note that although the analysis is focussed on finding eigenstates of the scalar constraint operator in the homogeneous-isotropic sector, the results have important impact on the spin-foam amplitudes we proposed in [42]. Our analysis in particular implies that a history of a state in $\mathcal{H}^j_{\text{cube}}$ is described by states in $\mathcal{H}^j_{\text{cube}}$. This means that in order $M$ of the expansion, the complexity of the problem gets roughly $24^{M-1}$ times smaller due to symmetry reduction.

## Acknowledgments

This work was supported by the National Science Centre, Poland Grant No. 2018/28/C/ST9/00157. The author would like to thank Mehdi Assanioussi for discussions. Numerical computations were done in the Świerk Computing Centre.

## ORCID iDs

Marcin Kisielowski  https://orcid.org/0000-0003-2875-0519